\documentclass[aps,twocolumn,showpacs,pra,twoside,amssymb,amsmath]{revtex4}
\usepackage{amssymb}
\usepackage{graphicx}
\usepackage{amsmath}
\usepackage{colordvi}
\usepackage{bbm}

\newcommand{\eins}{\ensuremath{\mathbbm 1}}

\begin{document}

\title{Entanglement detection via tighter local uncertainty relations}

\author{Cheng-Jie Zhang$^{1}$, Hyunchul Nha$^{2}$}
\author{Yong-Sheng Zhang$^{1}$}
\email{yshzhang@ustc.edu.cn}
\author{Guang-Can Guo$^{1}$}
\affiliation{$^1$Key Laboratory of Quantum Information, University
of Science and Technology of China, Hefei, Anhui 230026, People's
Republic of China\\
$^2$Department of Physics, Texas A \& M University at Qatar, Doha,
Qatar}

\begin{abstract}
We propose an entanglement criterion based on local
uncertainty relations (LURs) in a stronger form than the
original LUR criterion introduced in [H. F.
Hofmann and S. Takeuchi, Phys. Rev. A \textbf{68}, 032103 (2003)].
Using arbitrarily chosen operators $\{\hat{A}_{k}\}$ and
$\{\hat{B}_{k}\}$ of subsystems A and B, the tighter LUR criterion, which may be used not only for discrete
variables but also for continuous variables, can detect more entangled states
than the original criterion.
\end{abstract}


\maketitle

\section{Introduction}
Entangled states are usually recognized as essential resources in
quantum computation and communication \cite{nielsen}, and more and
more experimental realizations of entanglement sources have become
available \cite{experiment,experiment1,experiment2}.
However, there are still a number of important, yet open, problems concerning quantum entanglement.
In particular, the separability problem to determine both theoretically and experimentally
whether a given state is entangled or not is of crucial importance in quantum information science \cite{werner}.
In the past years, a great deal of efforts have been made to solve the separability problem
\cite{review1,Peres,LUR1,LUR2,CM,CM2,CCN1,CCN,permutation1,permutation2,dV,witness1,nEW,extension,LUR3,spin,Toth,nonlinear,Yu,detection2,genuine,Augusiak,concurrence,hierarchy,mintert,chen,Fei,Ma,Simon,Duan,Wolf,Giedke,Nha}.

There are many efficient methods proposed for entanglement
detection in both finite dimensional systems and continuous variable systems.
For example, in finite dimensional systems, there are
the partial transposition criterion
\cite{Peres}, local uncertainty relations (LUR) \cite{LUR1,LUR2},
covariance matrix criterion (CMC) \cite{CM,CM2}, the computable
cross-norm or realignment (CCNR) criterion \cite{CCN1,CCN}, the
permutation separability criteria \cite{permutation1,permutation2},
the criterion based on Bloch representations \cite{dV}, entanglement
witnesses \cite{witness1,nEW}, and Bell-type inequalities. On the
one hand, the partial transposition criterion is necessary and
sufficient for certain low dimensional systems, but it is known to
be only necessary for higher dimensions \cite{Peres}. The LUR
criterion provides only a necessary condition for arbitrary
dimensional systems, but it can detect many bound entangled states
where the partial transposition criterion fails \cite{LUR1}.
Moreover, it is shown in Ref. \cite{CM,CM2} that the LUR criterion
is equivalent to the symmetric CMC using orthogonal observables and
that the CCNR criterion together with its extension \cite{extension}
and the criterion based on Bloch representation are their
corollaries. The LUR, the symmetric CMC criteria, and their
corollaries are usually considered as complementary to the partial
transposition criterion. On the other hand, entanglement witnesses
and Bell-type inequalities are usually used for entanglement
detection in experiments. Recently, G\"uhne \textit{et al.} proposed
nonlinear witnesses to improve arbitrary linear witnesses, which is
strictly stronger than the original linear witnesses \cite{nEW}.
In continuous variable systems, Simon proposed a continuous variable
version of the partial transposition criterion in two-mode Gaussian
states \cite{Simon}. At the same time, Duan \textit{et al.} also introduced a criterion \cite{Duan}.
Both of these two criteria are necessary and sufficient conditions for
two-mode Gaussian states. Werner and Wolf improved Simon's result, and
they found bound entangled Gaussian states \cite{Wolf}. Furthermore, Giedke
\textit{et al.} provided a necessary and sufficient condition for
Gaussian states of bipartite systems of arbitrarily many modes \cite{Giedke}.

In this paper, we propose an entanglement criterion based on
LURs in a tighter form than the original LUR criterion.
Using arbitrarily chosen
operators $\{\hat{A}_{k}\}$ and $\{\hat{B}_{k}\}$ of subsystems A and
B, the stronger LUR criterion can generally detect more entangled states than the original LUR
criterion due to a newly added nonnegative term, similar to
the nonlinear witnesses.
Our tighter criterion can also be used both for discrete variables and for continuous variables.

The paper is organized as follows. In Sec. II we propose an
entanglement criterion based on the tighter LURs (TLURs) and illustrate its utility by an example of Horodecki $3\times3$ bound entangled states \cite{bound}.
In Sec. III the relationships between the TLUR criterion and other
entanglement criteria are discussed, and in Sec. IV,
a brief discussion and a summary of our results are given.

\section{Tighter local uncertainty relations}
In Ref. \cite{LUR1}, Hofmann and Takeuchi introduced an entanglement
criterion based on the local uncertainty relations. Consider the set
of local observables $\{\hat{A}_{k}\}_{k=1}^{N}$ and
$\{\hat{B}_{k}\}_{k=1}^{N}$ for subsystems A and B, respectively.
Suppose that the sum uncertainty relations have bounds for arbitrary local states as
\begin{eqnarray}
\sum_{k}\delta\hat{A}_{k}^{2}&\geq& U_{A},\label{suma}\\
\sum_{k}\delta\hat{B}_{k}^{2}&\geq& U_{B},\label{sumb}
\end{eqnarray}
where $U_{A}$ and $U_{B}$ are nonnegative values. Then, for separable
states, the following inequality holds \cite{LUR1},
\begin{equation}\label{LUR}
    \sum_{k}\delta(\hat{A}_{k}\otimes\eins+\eins\otimes\hat{B}_{k})_{\rho_{AB}}^{2}\geq
    U_{A}+U_{B}.
\end{equation}
It has been proven that the LUR criterion is equivalent to the
symmetric CMC using orthogonal observables, and that many other
criteria, such as the CCNR criterion together with its extension and
the criterion based on Bloch representation, are their corollaries
\cite{CM2}.  The LUR criterion is an efficient method to detect
bound entangled states, but is it possible to improve the LUR
criterion? Our idea comes from the nonlinear witnesses that improved
the linear witnesses \cite{nEW}. In the following, the LUR criterion
will be indeed developed in a tighter form to improve the power of
entanglement detection.

Before embarking on our criterion, a lemma will be given.
We again consider the sets of local observables
$\{\hat{A}_{k}\}_{k=1}^{N}$ and $\{\hat{B}_{k}\}_{k=1}^{N}$ for
subsystems A and B, which satisfy the bounds of the sum
uncertainty relations appearing in Eqs. (\ref{suma}) and (\ref{sumb}).
We first obtain the following lemma.

\textit{Lemma 1.} For bipartite separable states, the following
inequality must hold,
\begin{eqnarray}\label{}
  \sqrt{[\sum_{k}\delta(\hat{A}_{k})_{\rho_{A}}^{2}-U_{A}][\sum_{k}\delta(\hat{B}_{k})_{\rho_{B}}^{2}-U_{B}]}\nonumber\\
  \pm\sum_{k}(\langle\hat{A}_{k}\otimes\hat{B}_{k}\rangle-\langle\hat{A}_{k}\otimes\eins\rangle\langle\eins\otimes\hat{B}_{k}\rangle)\geq0.
\end{eqnarray}

\textit{Proof.--} The proof is given in the Appendix.
\hfill $\square$

\textit{Theorem 1.} (Tighter LURs) For
bipartite separable states, consider the sets of local observables
$\{\hat{A}_{k}\}_{k=1}^{N}$ and $\{\hat{B}_{k}\}_{k=1}^{N}$ for
subsystems A and B, respectively. If they satisfy the bounds of the sum
uncertainty relations in Eqs. (\ref{suma}) and (\ref{sumb}), then the
following inequality must hold,
\begin{eqnarray}
    &&\sum_{k}\delta(\hat{A}_{k}\otimes\eins+\eins\otimes\hat{B}_{k})_{\rho_{AB}}^{2}\geq U_{A}+U_{B}\nonumber\\
    &&
    +\Bigg[\sqrt{\sum_{k}\delta(\hat{A}_{k})_{\rho_{A}}^{2}-U_{A}}-\sqrt{\sum_{k}\delta(\hat{B}_{k})_{\rho_{B}}^{2}-U_{B}}\Bigg]^{2}.\label{NLUR}
\end{eqnarray}

\textit{Proof.--} Using Lemma 1, we can obtain that
\begin{eqnarray}
&&\sum_{k}\delta(\hat{A}_{k}\otimes\eins+\eins\otimes\hat{B}_{k})_{\rho_{AB}}^{2}\nonumber\\
&=&\sum_{k}\delta(\hat{A}_{k})_{\rho_{A}}^{2}+\sum_{k}\delta(\hat{B}_{k})_{\rho_{B}}^{2}\nonumber\\
&&+2\sum_{k}(\langle\hat{A}_{k}\otimes\hat{B}_{k}\rangle-\langle\hat{A}_{k}\otimes\eins\rangle\langle\eins\otimes\hat{B}_{k}\rangle)\nonumber\\
&\geq&\sum_{k}\delta(\hat{A}_{k})_{\rho_{A}}^{2}+\sum_{k}\delta(\hat{B}_{k})_{\rho_{B}}^{2}\nonumber\\
&&-2\sqrt{[\sum_{k}\delta(\hat{A}_{k})_{\rho_{A}}^{2}-U_{A}][\sum_{k}\delta(\hat{B}_{k})_{\rho_{B}}^{2}-U_{B}]}\nonumber\\
&=&U_{A}+U_{B}+M^{2},\nonumber
\end{eqnarray}
where
$M=\sqrt{\sum_{k}\delta(\hat{A}_{k})_{\rho_{A}}^{2}-U_{A}}-\sqrt{\sum_{k}\delta(\hat{B}_{k})_{\rho_{B}}^{2}-U_{B}}$.
\hfill $\square$

\textit{Remark.} It is worth noting that both Lemma 1 and Theorem 1
can be used for entanglement detection. Compared with the original
LUR criterion, the tighter criterion added a squared, thus nonnegative, term $M^{2}$.
Therefore, for a given set of observables
$\{\hat{A}_{k}\}_{k=1}^{N}$ and $\{\hat{B}_{k}\}_{k=1}^{N}$ at each
party, our criterion is stronger than the original LUR and can generally detect more entangled
states.

Actually, we can also prove that
$\sum_{k}\delta(\hat{A}_{k}\otimes\eins+\eins\otimes\hat{B}_{k})_{\rho_{AB}}^{2}\leq
U_{A}+U_{B}+(\sqrt{\sum_{k}\delta(\hat{A}_{k})_{\rho_{A}}^{2}-U_{A}}+\sqrt{\sum_{k}\delta(\hat{B}_{k})_{\rho_{B}}^{2}-U_{B}})^{2}$.
It is a dual inequality of Eq. (\ref{NLUR}).

\begin{figure}
\begin{center}
\includegraphics[scale=0.8]{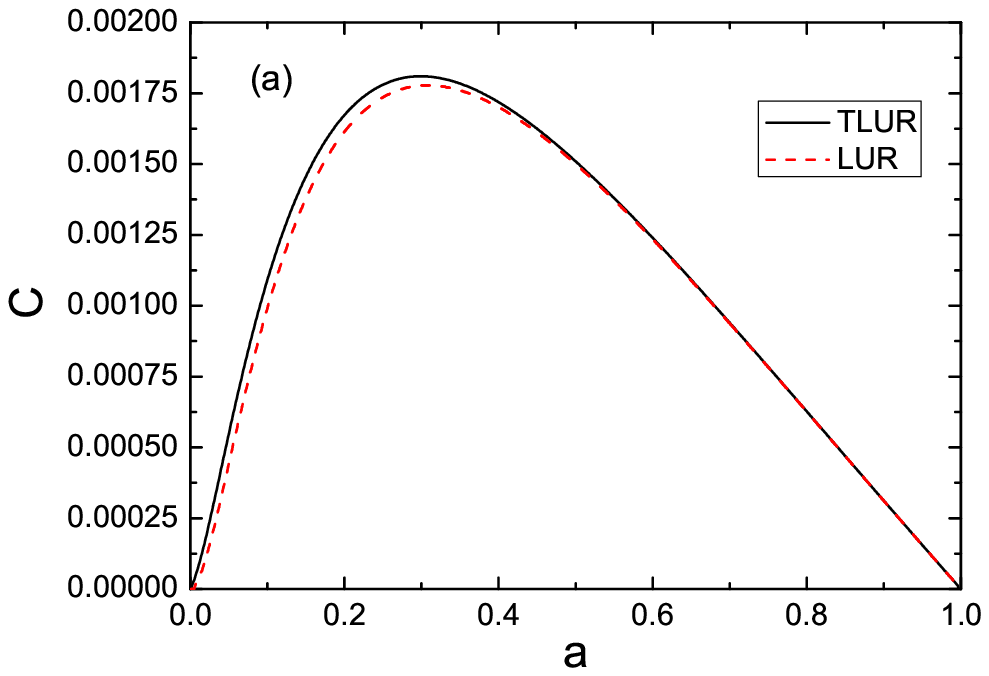}
\includegraphics[scale=0.8]{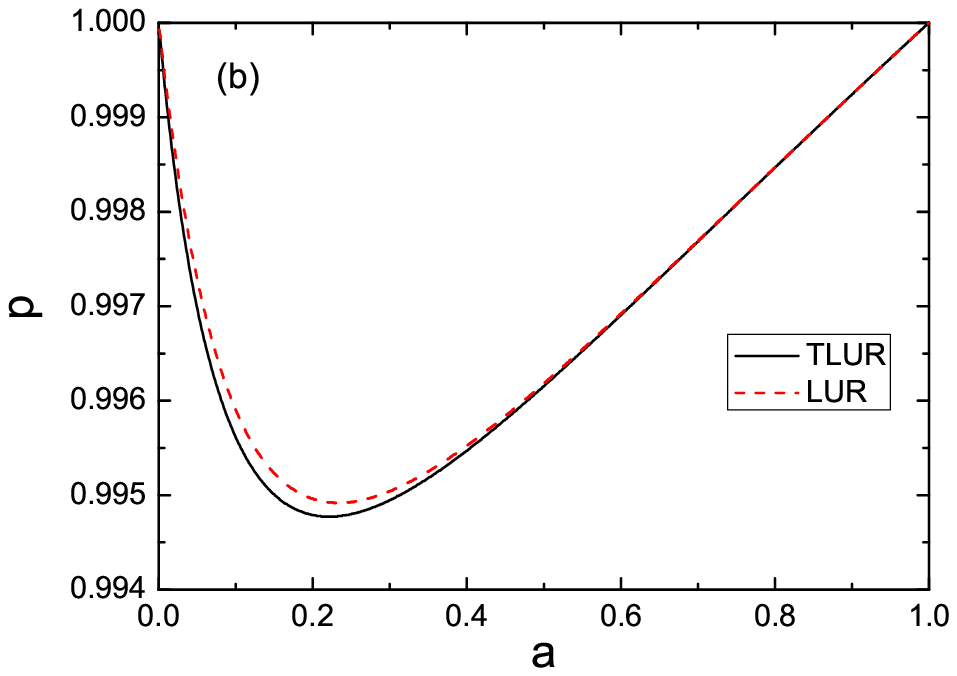}
\caption{(color online). (a) Describing the amount of entanglement for
Horodecki $3\times3$ bound entangled state via  $C_{TLUR}$ (solid
line) and $C_{LUR}$ (dashed line). $C_{TLUR}$ is always larger than
$C_{LUR}$. (b) Detecting the entanglement of Horodecki $3\times3$
bound entangled state with white noise. The regions above the curves
can be detected as entangled states by the TLUR criterion (solid
line) and the original LUR criterion (dashed line), respectively. }\label{1}
\end{center}
\end{figure}

\textit{Example 1.--} To compare with the original LUR criterion, we consider
the same example which has been used in Ref. \cite{LUR2}. P.
Horodecki introduced a $3\times3$ bound entangled state in Ref.
\cite{bound}, and its density matrix $\rho$ is real and symmetric as
\begin{eqnarray}
\label{eq:bound} \rho&=& \frac{a}{1+ 8 a} \big(
   \mid -1; 0 \rangle\langle -1; 0 \mid +
   \mid -1;+1 \rangle\langle -1;+1 \mid \nonumber \\
  && +\mid  0;-1 \rangle\langle  0;-1 \mid +
   \mid  0;+1 \rangle\langle  0;+1 \mid +
   \mid +1; 0 \rangle\langle +1; 0 \mid \big)
\nonumber \\ && + \frac{3 a}{1+ 8 a}
   \mid E_{\mbox{max}} \rangle\langle E_{\mbox{max}} \mid
                + \frac{1}{1+ 8 a}
   \mid \Pi \rangle\langle \Pi \mid,
\nonumber
\end{eqnarray}
where
\begin{eqnarray}\mid E_{\mbox{max}} \rangle =
\frac{1}{\sqrt{3}}\left( \mid -1;-1
\rangle + \mid 0;0 \rangle + \mid +1;+1\rangle \right) \nonumber
\end{eqnarray}
and
\begin{eqnarray}
\mid \Pi \rangle = \sqrt{\frac{1+a}{2}} \mid +1;-1
\rangle + \sqrt{\frac{1-a}{2}} \mid +1;+1 \rangle.\nonumber
\end{eqnarray}

The real parameter $a$ covers the range $0<a<1$. If one chooses the sets of
local observables $\{\hat{\lambda}_{k}(1)\}_{k=1}^{8}$ and
$\{\hat{\lambda}_{k}(2)\}_{k=1}^{8}$ introduced in Ref. \cite{LUR2},
the bound entangled state violates both the original LUR and the tighter LUR (TLUR) criterion.
On the other hand, if we define
$C_{TLUR}=1-[\sum_{k}\delta(\hat{A}_{k}\otimes\eins+\eins\otimes\hat{B}_{k})^{2}-M^{2}]/(U_{A}+U_{B})$
as Ref. \cite{LUR1} defined $C_{LUR}$, where $C_{TLUR}$ and $C_{LUR}$
provide quantitative estimates of the amount of entanglement
verified by the violation of LURs, it can be
discovered that $C_{TLUR}$ is always larger than $C_{LUR}$, which has
been demonstrated in Fig. 1(a). Furthermore, let us consider a mixture of
this state with white noise, $\rho(p)=p\rho+(1-p)\eins/9$. Taking the
TLUR in Eq.~(\ref{NLUR}) and the LUR in Eq.~(\ref{LUR}) with the Schmidt matrices of
$\rho(p)$ as the set of local observables \cite{nonlinear}, one
finds that more entangled states can be detected by the TLUR
criterion than by the LUR criterion. Detailed results are shown
in Fig. 1(b).

\section{Relation with other criteria}
In this section, we discuss the relation between Theorem 1 and some
other entanglement conditions that have been proposed in the past.
Actually, if we choose some special sets of local observables in
Theorem 1, there are several corollaries that can be obtained in a form
reduced to some other criteria or improved versions of them. One of
them is derived from local orthogonal observables (LOOs) $\hat{G}_{k}^{A}$
and $\hat{G}_{k}^{B}$ which are orthogonal bases of the observable
spaces $\mathcal{B}(\mathcal{H}_{A})$ and $\mathcal{B}(\mathcal{H}_{B})$
and satisfy
$\mathrm{Tr}(\hat{G}_{k}^{A}\hat{G}_{l}^{A})=\mathrm{Tr}(\hat{G}_{k}^{B}\hat{G}_{l}^{B})=\delta_{kl}$.

\textit{Corollary 1.} A stronger witness can be obtained from TLUR
using complete sets of LOOs as the set of local observables,
\begin{eqnarray}
1-\sum_{k}\langle \hat{G}_{k}^{A}\otimes
\hat{G}_{k}^{B}\rangle-\frac{1}{2}\sum_{k}\langle
\hat{G}_{k}^{A}\otimes\eins-\eins\otimes
\hat{G}_{k}^{B}\rangle^{2}\nonumber\\
-\frac{1}{2}\bigg(\sqrt{1-\mathrm{Tr}\rho_{A}^{2}}-\sqrt{1-\mathrm{Tr}\rho_{B}^{2}}\bigg)^{2}\geq0,\label{nonwitness}
\end{eqnarray}
Eq. (\ref{nonwitness}) holds for all bipartite separable states.

\textit{Proof.--} One can choose $\hat{A}_{k}=\hat{G}_{k}^{A}$,
$\hat{B}_{k}=-\hat{G}_{k}^{B}$, and use
$\sum_{k}\langle(\hat{G}_{k}^{A})^{2}\rangle=d_{A}$,
$\sum_{k}\langle\hat{G}_{k}^{A}\rangle^{2}=\mathrm{Tr}\rho_{A}^{2}$,
$\sum_{k}\langle(\hat{G}_{k}^{B})^{2}\rangle=d_{B}$, and
$\sum_{k}\langle\hat{G}_{k}^{B}\rangle^{2}=\mathrm{Tr}\rho_{B}^{2}$,
which have been shown in Ref. \cite{nonlinear}, to obtain Corollary
1. \hfill $\square$

\textit{Remark.} Corollary 1 is an improved version of the nonlinear
witness introduced in Ref. \cite{nonlinear}. Any entangled states,
which can be detected by the original nonlinear witness
\cite{nonlinear}, $1-\sum_{k}\langle \hat{G}_{k}^{A}\otimes
\hat{G}_{k}^{B}\rangle-\frac{1}{2}\sum_{k}\langle
\hat{G}_{k}^{A}\otimes\eins-\eins\otimes
\hat{G}_{k}^{B}\rangle^{2}\geq0$, can also be detected by Corollary
1; the converse is not true in general.

It is worth noticing that Corollary 1 can be easily realized in experiments, since
the left hand side of Eq. (\ref{nonwitness}) can be directly measured
($\sum_{k}\langle\hat{G}_{k}^{A}\rangle^{2}=\mathrm{Tr}\rho_{A}^{2}$,
$\sum_{k}\langle\hat{G}_{k}^{B}\rangle^{2}=\mathrm{Tr}\rho_{B}^{2}$). To show this,
we will provide a short example in the following.

\textit{Example 2.--} To compare with the original nonlinear witness, we
consider the same example which has been used in Ref. \cite{nonlinear}. Let
us consider a noisy singlet state of the form
\begin{equation}\label{}
    \rho(p)=p|\psi_{s}\rangle\langle\psi_{s}|+(1-p)\rho_{sep},
\end{equation}
where $|\psi_{s}\rangle=(|01\rangle-|10\rangle)/\sqrt{2}$ and
$\rho_{sep}=2/3|00\rangle\langle00|+1/3|01\rangle\langle01|$.
Actually, $\rho(p)$ is entangled for any $p>0$ \cite{nonlinear}.
Now we choose $\hat{G}_{k}^{A}$ and $\hat{G}_{k}^{B}$ as
\begin{eqnarray}
\{\hat{G}_{k}^{A}\}_{k=1}^{4}&=&\bigg\{-\frac{\sigma_{x}}{\sqrt{2}},-\frac{\sigma_{y}}{\sqrt{2}},-\frac{\sigma_{z}}{\sqrt{2}},\frac{\eins}{\sqrt{2}}\bigg\},\nonumber\\
\{\hat{G}_{k}^{B}\}_{k=1}^{4}&=&\bigg\{\frac{\sigma_{x}}{\sqrt{2}},\frac{\sigma_{y}}{\sqrt{2}},\frac{\sigma_{z}}{\sqrt{2}},\frac{\eins}{\sqrt{2}}\bigg\}.
\end{eqnarray}
It can be seen that $\rho(p)$ voilates the original nonlinear witness for all $p>0.25$ \cite{nonlinear}.
Using Eq. (\ref{nonwitness}) with these LOOs, one finds that $\rho(p)$
is entangled at least for $p>0.221$.

Besides Corollary 1, we can also obtain the conclusion that the CCNR
criterion, the criterion based on Bloch representations, and the
extension of CCNR criterion are the corollaries of Theorem 1. This
is because these three criteria are the corollaries of the symmetric
CMC criterion and that the symmetric CMC using orthogonal
observables is equivalent to the LUR criterion.

It is worth noticing that Theorem 1 can also be used for continuous
variables. If we choose $\{|a|\hat{x}_{1},|a|\hat{p}_{1}\}_{A}$ and
$\{\hat{x}_{2}/a,-\hat{p}_{2}/a\}_{B}$ as the sets of local
observables, and use
$\langle(\Delta\hat{x}_{j})^{2}\rangle+\langle(\Delta\hat{p}_{j})^{2}\rangle\geq|[\hat{x}_{j},\hat{p}_{j}]|=1$
for $j=1,2$, the following corollary will be obtained from
Theorem 1.

\textit{Corollary 2.} For continuous variable systems, define
$\hat{u}=|a|\hat{x}_{1}+\hat{x}_{2}/a$ and
$\hat{v}=|a|\hat{p}_{1}-\hat{p}_{2}/a$ with $\hat{x}_{j}$ and $\hat{p}_{j'}$ satisfying $[\hat{x}_{j},\hat{p}_{j'}]=i\delta_{jj'}$ ($j,j'=1,2$).
The inequality
\begin{eqnarray}
\langle(\Delta\hat{u})^{2}\rangle+\langle(\Delta\hat{v})^{2}\rangle\geq
a^{2}+\frac{1}{a^{2}}+M^{2}\label{corollary2}
\end{eqnarray}
holds for all separable states, where
$M=|a|\sqrt{\langle(\Delta\hat{x}_{1})^{2}\rangle+\langle(\Delta\hat{p}_{1})^{2}\rangle-1}-\sqrt{\langle(\Delta\hat{x}_{2})^{2}\rangle+\langle(\Delta\hat{p}_{2})^{2}\rangle-1}/|a|$.

\textit{Remark.} Corollary 2 is an improved version of the
entanglement criterion introduced in Ref. \cite{Duan}, which provided
an inequality
$\langle(\Delta\hat{u})^{2}\rangle+\langle(\Delta\hat{v})^{2}\rangle\geq
a^{2}+1/a^{2}$ for separable states. Since a squared nonnegative
term $M^{2}$ has been added in the right hand side of Eq.
(\ref{corollary2}), corollary 2 is strictly stronger than the
criterion shown in Ref. \cite{Duan}.

There is another interesting relation between the TLUR criterion and
the symmetric CMC using arbitrary observables. Notice that Refs.
\cite{CM,CM2} mainly discussed the symmetric CMC using orthogonal
observables and concluded that the LUR criterion is equivalent to
the symmetric CMC using \textit{orthogonal} observables. Interestingly, if
\textit{arbitrary} local observables $\{\hat{A}_{k}\}$ and
$\{\hat{B}_{k}\}$ are used, the TLUR
can be obtained from the symmetric CMC \cite{otfried}.

\textit{Proposition 1.} The TLUR criterion is a corollary of the
symmetric CMC using \textit{arbitrary} local observables \cite{otfried}.

\textit{Proof.---} Using Eq. (43) of Ref. \cite{CM2}
$\|C\|^{2}_{Tr}\leq\|A-\kappa_{A}\|_{Tr}\|B-\kappa_{B}\|_{Tr}$ where
$\|\cdot\|_{Tr}$ is the trace norm (i.e., the sum of the singular
values), $\kappa_{A}=\sum_{i}p_{i}\gamma^{S}(\rho_{i}^{A})$,
$\kappa_{B}=\sum_{i}p_{i}\gamma^{S}(\rho_{i}^{B})$, and $\gamma^{S}$
stands for the symmetric covariance matrix, one can obtain that
$[\sum_{k}(\langle\hat{A}_{k}\otimes\hat{B}_{k}\rangle-\langle\hat{A}_{k}\otimes\eins\rangle\langle\eins\otimes\hat{B}_{k}\rangle)]^2=(\mathrm{Tr}C)^{2}\leq\|C\|^{2}_{Tr}\leq\|A-\kappa_{A}\|_{Tr}\|B-\kappa_{B}\|_{Tr}
=(\mathrm{Tr}A-\mathrm{Tr}\kappa_{A})(\mathrm{Tr}B-\mathrm{Tr}\kappa_{B})=[\sum_{k}\delta(\hat{A}_{k})_{\rho_{A}}^{2}-\sum_{i}p_{i}\sum_{k}\delta(\hat{A}_{k})_{\rho^{A}_{i}}^{2}][\sum_{k}\delta(\hat{B}_{k})_{\rho_{B}}^{2}-\sum_{i}p_{i}\sum_{k}\delta(\hat{B}_{k})_{\rho^{B}_{i}}^{2}]
\leq[\sum_{k}\delta(\hat{A}_{k})_{\rho_{A}}^{2}-U_{A}][\sum_{k}\delta(\hat{B}_{k})_{\rho_{B}}^{2}-U_{B}]$,
where $\sum_{k}\delta(\hat{A}_{k})_{\rho^{A}_{i}}^{2}\geq U_{A}$ and
$\sum_{k}\delta(\hat{B}_{k})_{\rho^{B}_{i}}^{2}\geq U_{B}$ have been
used. Therefore, Lemma 1 and Theorem 1 can be obtained from the
symmetric CMC using arbitrary observables. \hfill $\square$

\textit{Remark.} Refs. \cite{CM,CM2} show that the LUR criterion using
arbitrary observables is equivalent to the symmetric CMC using \textit{orthogonal}
observables. Obviously, the symmetric CMC using \textit{orthogonal} observables is
a corollary of the symmetric CMC using \textit{arbitrary} observables. From Proposition 1,
TLUR criterion is also a corollary of the symmetric CMC using \textit{arbitrary} observables.
However, whether the TLUR criterion is equivalent to the LUR criterion (the symmetric CMC using
\textit{orthogonal} observables) is unknown.


\section{Discussion and conclusion}
There are still several questions about the TLUR. First, Theorem 1
and the original LUR criterion are considered for bipartite systems.
Is it possible to generalize them to multipartite systems? Second,
we have shown that Theorem 1 is stronger than the LUR when the set of
local observables is chosen. However, if one chooses all possible
sets of local observables, is Theorem 1 still stronger than or just
equivalent to the LUR criterion? Finally, for discrete variable
systems, Theorem 1 can be used for detecting bound entangled states.
Is it then possible to detect bound entangled states for continuous
variables? These questions are interesting and worth for further
research.

In summary, we have proposed an entanglement criterion based on
the TLUR, which can be viewed as an extension of the original LUR
criterion. Using arbitrarily chosen operators $\{\hat{A}_{k}\}$ and
$\{\hat{B}_{k}\}$ of subsystems A and B, the TLUR criterion,
which may be used not only for discrete variables but also for continuous variables, can detect
more entangled states than the LUR criterion since a nonnegative term
has been added, similar to the nonlinear witnesses.

\section*{ACKNOWLEDGMENTS}
We would like to thank Otfried G\"uhne for helpful discussions, and
anonymous referee for valuable suggestions. This work was funded by the National
Fundamental Research Program (Grant No. 2006CB921900), the National
Natural Science Foundation of China (Grants No. 10674127, No. 60621064 and No. 10974192),
the Innovation Funds from the Chinese Academy of Sciences, and the K.C. Wong
Foundation. HN is supported by an NPRP grant 1-7-7-6 from Qatar National Research
Funds.

\section*{APPENDIX} Here we prove Lemma 1. The density matrix for bipartite separable states
can be expressed as
$\rho_{AB}=\sum_{i}p_{i}\rho_{i}^{A}\otimes\rho_{i}^{B}$. Notice
that the lemma is equivalent to
\begin{eqnarray}
[\sum_{k}\delta(\hat{A}_{k})_{\rho_{A}}^{2}-U_{A}][\sum_{k}\delta(\hat{B}_{k})_{\rho_{B}}^{2}-U_{B}]\nonumber\\
\geq[\sum_{k}(\langle\hat{A}_{k}\otimes\hat{B}_{k}\rangle-\langle\hat{A}_{k}\otimes\eins\rangle\langle\eins\otimes\hat{B}_{k}\rangle)]^{2}.\label{lemma}
\end{eqnarray}
The right hand side (RHS) and the left hand side (LHS) of Eq.
(\ref{lemma}) can be written as
\begin{eqnarray}
&&\mathrm{RHS}\nonumber\\
&=&\{\sum_{k}[\sum_{i}p_{i}\langle\hat{A}_{k}\rangle_{i}\langle\hat{B}_{k}\rangle_{i}-(\sum_{i}p_{i}\langle\hat{A}_{k}\rangle_{i})(\sum_{i'}p_{i'}\langle\hat{B}_{k}\rangle_{i'})]\}^{2}\nonumber\\
&=&\{\sum_{k}\sum_{i}p_{i}\langle(\hat{A}_{k}-\sum_{i'}p_{i'}\langle\hat{A}_{k}\rangle_{i'})\rangle_{i}\langle(\hat{B}_{k}-\sum_{j'}p_{j'}\langle\hat{B}_{k}\rangle_{j'})\rangle_{i}\}^{2}\nonumber\\
&\equiv&\{\sum_{k}\sum_{i}p_{i}\langle\mathfrak{A}_{k}\rangle_{i}\langle\mathfrak{B}_{k}\rangle_{i}\}^{2},\nonumber
\end{eqnarray}
where we have defined
$\langle\hat{A}_{k}\rangle_{i}=\langle\hat{A}_{k}\rangle_{\rho_{i}^{A}}$,
$\langle\hat{B}_{k}\rangle_{i}=\langle\hat{B}_{k}\rangle_{\rho_{i}^{B}}$,
$\mathfrak{A}_{k}=\hat{A}_{k}-\sum_{i'}p_{i'}\langle\hat{A}_{k}\rangle_{i'}$
and
$\mathfrak{B}_{k}=\hat{B}_{k}-\sum_{j'}p_{j'}\langle\hat{B}_{k}\rangle_{j'}$
for convenience. Therefore,
\begin{eqnarray}
&&\mathrm{LHS}\nonumber\\
&=&(\sum_{k}\sum_{i}p_{i}\langle(\mathfrak{A}_{k})^{2}\rangle_{i}-U_{A})(\sum_{k}\sum_{i}p_{i}\langle(\mathfrak{B}_{k})^{2}\rangle_{i}-U_{B})\nonumber\\
&\geq&(\sum_{k}\sum_{i}p_{i}\langle\mathfrak{A}_{k}\rangle_{i}^{2})(\sum_{k}\sum_{i}p_{i}\langle\mathfrak{B}_{k}\rangle_{i}^{2}),\nonumber
\end{eqnarray}
where we have used
$\sum_{k}\sum_{i}p_{i}(\langle(\mathfrak{A}_{k})^{2}\rangle_{i}-\langle\mathfrak{A}_{k}\rangle_{i}^{2})=\sum_{k}\sum_{i}p_{i}(\langle(A_{k})^{2}\rangle_{i}-\langle
A_{k}\rangle_{i}^{2})\geq U_{A}$ and
$\sum_{k}\sum_{i}p_{i}(\langle(\mathfrak{B}_{k})^{2}\rangle_{i}-\langle\mathfrak{B}_{k}\rangle_{i}^{2})=\sum_{k}\sum_{i}p_{i}(\langle(B_{k})^{2}\rangle_{i}-\langle
B_{k}\rangle_{i}^{2})\geq U_{B}$.

With the help of the Cauchy-Schwarz inequality, it can be obtained
that
\begin{eqnarray}
\mathrm{LHS}
&\geq&(\sum_{k}\sum_{i}p_{i}\langle\mathfrak{A}_{k}\rangle_{i}^{2})(\sum_{k}\sum_{i}p_{i}\langle\mathfrak{B}_{k}\rangle_{i}^{2})\nonumber\\
&\geq&(\sum_{k}\sum_{i}p_{i}\langle\mathfrak{A}_{k}\rangle_{i}\langle\mathfrak{B}_{k}\rangle_{i})^{2}\nonumber\\
&=&\mathrm{RHS}.\nonumber
\end{eqnarray}
Therefore, Lemma 1 has been proved. \hfill $\square$


\begin{thebibliography}{99}


\bibitem{nielsen} M. A. Nielsen and I. L. Chuang, \textit{Quantum
Computation and Quantum Information} (Cambridge University Press,
Cambridge, 2000).

\bibitem{experiment} D. Leibfried, E. Knill, S. Seidelin, J. Britton, R. B. Blakestad, J. Chiaverini, D. B. Hume, W. M. Itano,
J. D. Jost, C. Langer, R. Ozeri, R. Reichle, and D. J. Wineland,
Nature \textbf{438}, 639 (2005).

\bibitem{experiment1} H. H\"affner, W. H\"ansel, C. F. Roos, J. Benhelm, D. Chek-al-kar, M. Chwalla, T. K\"orber, U. D. Rapol,
M. Riebe, P. O. Schmidt, C. Becher, O. G\"uhne, W. D\"ur, and R.
Blatt, Nature \textbf{438}, 643 (2005).

\bibitem{experiment2} C.-Y. Lu, X.-Q. Zhou, O. G\"uhne, W.-B. Gao, J. Zhang, Z.-S. Yuan,
A. Goebel, T. Yang and J.-W. Pan, Nature Phys. \textbf{3}, 91
(2007).

\bibitem{werner} R. F. Werner, Phys. Rev. A \textbf{40}, 4277 (1989).



\bibitem{review1} D. Bru\ss, J. Math. Phys. \textbf{43}, 4237 (2002);
 M. B. Plenio, S. Virmani, Quantum Inf. Comput. \textbf{7}, 1
(2007); R. Horodecki, P. Horodecki, M. Horodecki, and K. Horodecki,
Rev. Mod. Phys. \textbf{81}, 865 (2009); O. G\"uhne and G. T\'{o}th,
Phys. Rep. \textbf{474}, 1 (2009).

\bibitem{Peres} A. Peres, Phys. Rev. Lett. \textbf{77}, 1413 (1996);
M. Horodecki, P. Horodecki, and R. Horodecki, Phys. Lett. A
\textbf{223}, 1 (1996).

\bibitem{LUR1} H. F. Hofmann and S. Takeuchi, Phys. Rev. A
\textbf{68}, 032103 (2003).


\bibitem{LUR2} H. F. Hofmann, Phys. Rev. A \textbf{68}, 034307
(2003).

\bibitem{CM} O. G\"uhne, P. Hyllus, O. Gittsovich, and J. Eisert, Phys. Rev.
Lett. \textbf{99}, 130504 (2007).

\bibitem{CM2} O. Gittsovich, O. G\"uhne, P. Hyllus, and J. Eisert, Phys. Rev. A
\textbf{78}, 052319 (2008).

\bibitem{CCN1} O. Rudolph, arXiv:quant-ph/0202121.

\bibitem{CCN} K. Chen and L.-A. Wu, Quantum Inf. Comput. \textbf{3}, 193 (2003).

\bibitem{permutation1} M. Horodecki, P. Horodecki, and R. Horodecki,
Open Syst. Inf. Dyn. \textbf{13}, 103 (2006).

\bibitem{permutation2} H. Fan, arXiv:quant-ph/0210168; P. Wocjan, M. Horodecki, Open Syst. Inf. Dyn. \textbf{12}, 331 (2005);  L. Clarisse, P.
Wocjan, Quantum Inf. Comput. \textbf{6}, 277 (2006).

\bibitem{dV} J.I. de Vicente, Quantum Inf. Comput. \textbf{7}, 624
(2007); J.I. de Vicente, J. Phys. A \textbf{41}, 065309 (2008).


\bibitem{witness1} B. Terhal, Phys. Lett. A \textbf{271}, 319 (2000);
G. T\'oth and O. G\"uhne, Phys. Rev. Lett. \textbf{94}, 060501
(2005); F. A. Bovino, G. Castagnoli, A. Ekert, P. Horodecki, C. M.
Alves and A. V. Sergienko, Phys. Rev. Lett. \textbf{95}, 240407
(2005); O. G\"uhne, G. T\'oth, P. Hyllus, and H. J. Briegel, {\it
ibid.} \textbf{95}, 120405 (2005); F. Mintert, Phys. Rev. A
\textbf{75}, 052302 (2007); R. Augusiak, M. Demianowicz, P.
Horodecki, {\it ibid.} \textbf{77}, 030301(R) (2008).


\bibitem{nEW} O. G\"uhne and N. L\"utkenhaus, Phys. Rev. Lett.
\textbf{96}, 170502 (2006);

\bibitem{extension} C.-J. Zhang, Y.-S. Zhang, S. Zhang, and G.-C.
Guo, Phys. Rev. A \textbf{77}, 060301(R) (2008).

\bibitem{LUR3} O. G\"uhne, Phys. Rev. Lett. \textbf{92}, 117903 (2004).

\bibitem{spin} G. T\'oth, C. Knapp, O. G\"uhne, and H. J. Briegel,
Phys. Rev. Lett. \textbf{99}, 250405 (2007); G. T\'oth, C. Knapp, O.
G\"uhne, and H. J. Briegel, Phys. Rev. A \textbf{79}, 042334 (2009).

\bibitem{Toth} G. T\'oth,and O. G\"uhne, Phys. Rev. Lett. \textbf{102}, 170503 (2009); G. T\'oth, W. Wieczorek, R. Krischek,
N. Kiesel, P. Michelberger, and H. Weinfurter, New J. Phys.
\textbf{11}, 083002 (2009).



\bibitem{nonlinear} O. G\"uhne, M. Mechler, G. T\'oth, and P. Adam, Phys. Rev. A
\textbf{74}, 010301(R) (2006).


\bibitem{Yu} S. Yu and N.-L. Liu, Phys. Rev. Lett. \textbf{95},
150504 (2005).

\bibitem{detection2} P. Aniello and C. Lupo, J. Phys. A: Math. Theor. \textbf{41}, 355303 (2008); C. Lupo, P. Aniello, and A.
Scardicchio, {\it ibid.} \textbf{41}, 415301 (2008).

\bibitem{genuine} O. G\"uhne amd M. Seevinck, arXiv:0905.1349.

\bibitem{Augusiak} R. Augusiak and J. Stasi\'nska, New J. Phys.
\textbf{11}, 053018 (2009).



\bibitem{concurrence} W. K. Wootters, Phys. Rev. Lett. \textbf{80},
2245 (1998); A. Uhlmann, Phys. Rev. A \textbf{62}, 032307 (2000); P.
Rungta, V. Bu\v zek, C. M. Caves, M. Hillery and G. J. Milburn,
Phys. Rev. A \textbf{64}, 042315 (2001).

\bibitem{hierarchy} H. Fan, K. Matsumoto and H. Imai, J. Phys. A \textbf{36},
4151 (2003); H. Fan, V. Korepin, and V. Roychowdhury, Phys. Rev.
Lett. \textbf{93}, 227203 (2004).

\bibitem{mintert} F. Mintert, M. Ku\'s and A. Buchleitner, Phys.
Rev. Lett. \textbf{92}, 167902 (2004); H. P. Breuer, J. Phys. A
\textbf{39}, 11847 (2006); J.I. de Vicente, Phys. Rev. A
\textbf{75}, 052320 (2007).

\bibitem{chen} K. Chen, S. Albeverio and S.-M. Fei, Phys. Rev. Lett.
\textbf{95}, 040504 (2005); K. Chen, S. Albeverio and S.-M. Fei,
{\it ibid.} 95, 210501 (2005).

\bibitem{Fei} M. Li, S.-M. Fei, and Z.-X. Wang, J. Phys. A \textbf{41},
202002 (2008); M.-J. Zhao, Z.-X. Wang, and S.-M. Fei, Rep. Math. Phys.
\textbf{63}, 409 (2009); S.-M. Fei, M.-J. Zhao, K. Chen, and Z.-X. Wang,
Phys. Rev. A \textbf{80}, 032320 (2009).

\bibitem{Ma} Z.-H. Ma, F.-L. Zhang, D.-L. Deng, and J.-L. Chen,
Phys. Lett. A \textbf{373}, 1616 (2009); Z.-H. Ma, F.-L. Zhang,
and J.-L. Chen, Phys. Rev. A \textbf{78}, 064305 (2008).

\bibitem{Simon} R. Simon, Phys. Rev. Lett. \textbf{84}, 2726 (2000).

\bibitem{Duan} L.-M. Duan, G. Giedke, J. I. Cirac, and P. Zoller,
Phys. Rev. Lett. \textbf{84}, 2722 (2000).

\bibitem{Wolf} R. F. Werner and M. M. Wolf, Phys. Rev. Lett. \textbf{86},
3658 (2001).

\bibitem{Giedke} G. Giedke, B. Kraus, M. Lewenstein, and J. I. Cirac,
Phys. Rev. Lett. \textbf{87}, 167904 (2001).

\bibitem{Nha} H. Nha, and M. S. Zubairy, Phys. Rev. Lett.
\textbf{101}, 130402 (2008); Q. Sun, H. Nha, and M. S. Zubairy,
Phys. Rev. A \textbf{80}, 020101(R) (2009).



\bibitem{bound} P. Horodecki, Phys. Lett. A \textbf{232}, 333 (1997).

\bibitem{otfried} O. G\"uhne, Private communication.




\end{thebibliography}
\end{document}